\documentclass[aps,prl,twocolumn,groupedaddress,showpacs,nofootinbib]{revtex4-1}
\usepackage{graphicx,hyperref}
\newcommand{\apjl}{ApJL}
\newcommand{\mnras}{MNRAS}

\begin{document}

%=====================================================================
\title{Tests of Local Lorentz Invariance Violation of Gravity
  \\ in the Standard-Model Extension with Pulsars}
\author{Lijing Shao}
\email{lshao@pku.edu.cn}
\affiliation{Max-Planck-Institut f\"ur Radioastronomie, Auf dem
  H\"ugel 69, D-53121 Bonn, Germany \\ School of Physics, Peking
  University, Beijing 100871, China}  
%\date{\today}
%=====================================================================
\begin{abstract}
  Standard-model extension (SME) is an effective field theory
  introducing all possible Lorentz-violating (LV) operators to the
  standard model (SM) and general relativity (GR). In the pure-gravity
  sector of minimal SME (mSME), nine coefficients describe dominant
  observable deviations from GR. We systematically implemented
  twenty-seven tests from thirteen pulsar systems to tightly constrain
  eight linear combinations of these coefficients with extensive Monte
  Carlo simulations. It constitutes the first detailed and systematic
  test of the pure-gravity sector of mSME with the state-of-the-art
  pulsar observations. No deviation from GR was detected. The limits
  of LV coefficients are expressed in the canonical Sun-centered
  celestial-equatorial frame for convenience of further studies. They
  are all improved by significant factors of tens to hundreds with
  existing ones. As a consequence, Einstein's equivalence principle is
  verified substantially further by pulsar experiments in terms of
  local Lorentz invariance in gravity. 
\end{abstract}
%=====================================================================
\maketitle

%=====================================================================
\paragraph{Introduction.\label{sec:intro}}
%=====================================================================

Einstein's general relativity (GR) has passed all experimental
scrutinies for almost one hundred years with flying
colors~\cite{will93,will06}. However, besides current puzzles on the
nature of dark matter and dark energy, there exist difficulties to
combine GR and the standard model (SM) into a unified theory.  A full
theory of quantum gravity (QG) may settle the issues, but such a
theory is still missing. Although one expects that the full QG theory
will {\it reduce to} or {\it emerge as} GR for the gravity interaction
in our accessible energy regime, it may still leave {\it relic
  effects} as {\it QG
  windows}~\cite{ks89a,ks89b,ck97,ck98,gp99,kli00,jm01,km02,kos04,bk05,bk06,hor09,kt11,kr11,ame13,lib13}. They
are widely predicted by string theory, loop quantum gravity, and
noncommutative field theories, wherein Lorentz-violating (LV) effects
are well-known examples. Because of the importance of gravity, as a
fundamental interaction, and Lorentz symmetry, as a fundamental
property of spacetime, persistent efforts to probe possible LV
deviations from GR are well justified and readily
desired~\cite{ks89a,ks89b,ck97,ck98,gp99,jm01,km02,kos04,bk05,bk06,hor09,kt11,kr11,ame13,lib13,nac69,cn83,kli00}.

Recently, there is a strong belief that GR and SM are effective field
theories (EFTs) of the underlying full theory~\cite{wei09}. With the
fact that GR and SM have passed all tests up to now, one would expect
that LV deviations, if any, would be suppressed by a high energy scale.
Hence it is natural to include extra terms in the context of
EFTs. Standard-model extension (SME) is constructed as a convenient
experimentally working framework to probe all possible LV deviations
in the spirit of
EFTs~\cite{ks89a,ks89b,ck97,ck98,km02,kos04,bk05,bk06,kt11,kr11}.

In this Letter, for the first time, we systematically used various
state-of-the-art pulsar observations to constrain LV effects in the
pure-gravity sector of minimal SME (mSME)~\cite{bk06}. We improved all
limits over previous ones~\cite{bcs07,mch+08,cch+09} by substantial
factors of tens to hundreds. Einstein's equivalence principle is thus
verified further in terms of local Lorentz invariance in
gravity. Light speed $c=1$ is adopted throughout.

%=====================================================================
\paragraph{Pure-gravity sector of mSME.\label{sec:sme}} 
%=====================================================================

In SME, a general Lagrangian including gravity in Riemann-Cartan
spacetime has the structure, ${\cal L} = {\cal L}_{\rm LI} + {\cal
  L}_{\rm LV}$, where ${\cal L}_{\rm LI}$ and ${\cal L}_{\rm LV}$ are
Lorentz-invariant (LI) and LV terms respectively~\cite{kos04}. We
focus on the limit of Riemannian spacetime and the pure-gravity sector
with LV operators of only mass dimension four or less (the so-called
mSME). Then ${\cal L}_{\rm LI} = \sqrt{-g} (R-2\Lambda)/16\pi G$ is
the usual Einstein-Hilbert action of GR, with $g$ the determinant of
the metric, $R$ the Ricci scalar, and $\Lambda$ the cosmological
constant that is set to zero for localized systems. The LV Lagrangian
at leading order reads~\cite{kos04,bk06},
%---------------------------------------------------------------------
\begin{equation}
  {\cal L}_{\rm LV} = \frac{\sqrt{-g}}{16\pi G} \left( - u R +
  s^{\mu\nu} R_{\mu\nu}^{\rm T} + t^{\kappa\lambda\mu\nu}
  C_{\kappa\lambda\mu\nu} \right) \,, \label{eq:S_LV}
\end{equation}
%---------------------------------------------------------------------
where $R^{\rm T}_{\mu\nu}$ is the trace-free Ricci tensor and
$C_{\kappa\lambda\mu\nu}$ is the Weyl conformal tensor.  The LV
fields, $u$, $s^{\mu\nu}$, and $t^{\kappa\lambda\mu\nu}$, violate both
the {\it particle} local Lorentz invariance and the diffeomorphism,
while the {\it observer} local Lorentz invariance is
preserved. Because $s^{\mu\nu}$ and $t^{\kappa\lambda\mu\nu}$ inherit
the symmetries of $R^{\rm T}_{\mu\nu}$ and $C_{\kappa\lambda\mu\nu}$
respectively, there exist in total twenty independent LV
coefficients~\cite{bk06}. It is interesting to note that, if the
symmetry breaking is spontaneous, SME coefficients arise from the
underlying dynamics, so they must be regarded as dynamical fields. In
contrast, if the breaking is explicit, the fields originate as
prescribed spacetime functions that play no dynamical r\^ole.  The
geometry of Riemann-Cartan spacetime prohibits the
latter~\cite{kos04,bk05}. Hence, only spontaneous breaking is
considered here. After properly accounting for Nambu-Goldstone modes
and adopting several plausible assumptions, Bailey and Kosteleck\'y
found that nine components of a trace-free matrix $\bar s^{\mu\nu}$,
which are the (rescaled) vacuum expectation values of $s^{\mu\nu}$,
describe dominant observable effects; see~\cite{bk06} for details.

%=====================================================================
\paragraph{Coordinate systems.\label{sec:coord}}
%=====================================================================

%---------------------------------------------------------------------
\begin{figure}
  \includegraphics[width=6cm]{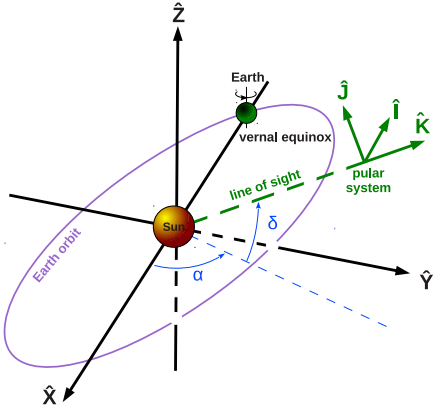}
  \caption{The canonical spatial reference frame for SME is the
    Sun-centered celestial-equatorial frame $(\hat{\bf X}, \hat{\bf
      Y}, \hat{\bf Z})$, with $\hat{\bf X}$ pointing from the Earth to
    the Sun at vernal equinox, $\hat{\bf Z}$ along the rotating axis
    of the Earth, and $\hat{\bf Y} \equiv \hat{\bf Z} \times \hat{\bf
      X}$~\cite{km02}.  The frame $(\hat{\bf I}, \hat{\bf J}, \hat{\bf
      K})$ is comoving with the pulsar system, with $\hat{\bf K}$
    pointing along the line of sight to the pulsar, while $(\hat{\bf
      I}, \hat{\bf J})$ constitutes the sky plane with $\hat{\bf I}$
    to east, and $\hat{\bf J}$ to north~\cite{dt92}. Besides a boost
    of ${\cal O}(10^{-3})$, these two frames are related by matrices
    ${\cal R}^{(\alpha)}$ and ${\cal R}^{(\delta)}$~\cite{sm}.
    $(\hat{\bf I}, \hat{\bf J}, \hat{\bf K})$ is denoted as
    $(\vec{e}_1, \vec{e}_2, \vec{e}_3)$
    in~\cite{bk06}.\label{fig:solar_geometry}}
\end{figure}
%---------------------------------------------------------------------

\begin{figure}
  \includegraphics[width=5.5cm]{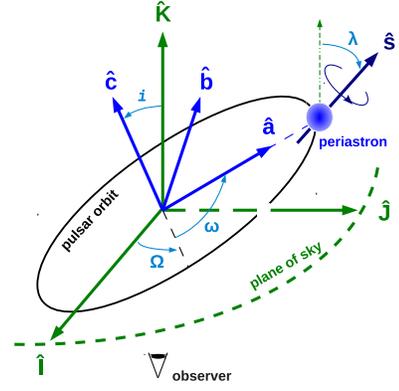}
  \caption{The spatial frame $(\hat{\bf a}, \hat{\bf b}, \hat{\bf c})$
    is centered at the pulsar system with $\hat{\bf a}$ pointing from
    the center of mass to the periastron, $\hat{\bf c}$ along the
    orbital angular momentum, and $\hat{\bf b} \equiv \hat{\bf c}
    \times \hat{\bf a}$. It is related to $(\hat{\bf I}, \hat{\bf J},
    \hat{\bf K})$ frame through rotation matrices ${\cal
      R}^{(\Omega)}$, ${\cal R}^{(i)}$, and ${\cal
      R}^{(\omega)}$~\cite{sm}.  The spin direction $\hat{\bf S}$ has
    a polar angle $\lambda$ and an azimuthal angle $\eta$ (not shown)
    in $(\hat{\bf I}, \hat{\bf J}, \hat{\bf K})$ frame. $(\hat{\bf a},
    \hat{\bf b}, \hat{\bf c})$ is denoted as $(\vec{P}, \vec{Q},
    \vec{k})$ in~\cite{bk06}.\label{fig:pulsar_geometry}}
\end{figure}

The tensorial background $\bar s^{\mu\nu}$ is {\it observer} LI, while
{\it particle} LV. Therefore, to probe the magnitudes of $\bar
s^{\mu\nu}$, one should explicitly point out the observer coordinate
system in use.  In the context of post-Newtonian gravity of SME, the
standard frame is an asymptotically inertial frame, $({\rm T},
\hat{\bf X},\hat{\bf Y},\hat{\bf Z})$, that is comoving with the Solar
System~\cite{bk06} (see Fig.~\ref{fig:solar_geometry}).  For a pulsar
binary, the most convenient frame, $(t,\hat{\bf a},\hat{\bf
  b},\hat{\bf c})$, is defined by the orbit (see
Fig.~\ref{fig:pulsar_geometry}).  To relate two frames, in general a
Lorentz transformation is required~\cite{bk06}. It consists of a
spatial rotation, ${\cal R}$, to align $(\hat{\bf a},\hat{\bf
  b},\hat{\bf c})$ and $(\hat{\bf X},\hat{\bf Y},\hat{\bf Z})$, and a
boost characterized by the relative velocity of the pulsar system with
respect to the Solar system. Typically, its magnitude equals to ${\cal
  O}(10^2\,{\rm km/s})$, which implies a boost of ${\cal O}(10^{-3})$
that is neglected here.  We are left with a pure spatial rotation
${\cal R}$. It can be obtained with the help of an intermediate
coordinate system, $(\hat{\bf I}, \hat{\bf J}, \hat{\bf K})$ (see
Figs.~\ref{fig:solar_geometry}--\ref{fig:pulsar_geometry}). It
involves five simple steps to transform from $(\hat{\bf a},\hat{\bf
  b},\hat{\bf c})$ to $(\hat{\bf X},\hat{\bf Y},\hat{\bf Z})$,
characterized by $\alpha$ (right ascension), $\delta$ (declination),
$\Omega$ (longitude of ascending node), $i$ (orbital inclination), and
$\omega$ (longitude of periastron). The full rotation ${\cal R} =
{\cal R}^{(\omega)} {\cal R}^{(i)} {\cal R}^{(\Omega)} {\cal
  R}^{(\delta)} {\cal R}^{(\alpha)}$~\cite{sm}. The transformations of
$\bar s^{\mu\nu}$ are $\bar s^{tt} \doteq \bar s^{\rm TT}$, $\bar
s^{AB} \doteq {\cal R}^{A}_{~x} {\cal R}^{B}_{~y} \bar s^{xy}$, and
$\bar s^{tA} \doteq {\cal R}^{A}_{~x} \bar s^{{\rm T}x}$, where
$A,B=a,b,c$ and $x,y={\rm X,Y,Z}$.

%---------------------------------------------------------------------
%---------------------------------------------------------------------

%=====================================================================
\paragraph{Spin precession of solitary pulsars.\label{sec:solitary}} 
%=====================================================================

Following the derivations in~\cite{nor87}, one gets an extra
precession, ${\bf \Omega}^{\rm prec}$, for an isolated spinning body
in internal equilibrium~\cite{bk06}. The precession rate is
$\Omega^{\rm prec}_k = \pi \bar s^{jk} \hat{S}^j/P$, where $P$ is the
spin period, and $\hat{\bf S}$ is the unit vector pointing along the
spin direction. Because of their small spin periods, millisecond
pulsars (MSPs) are ideal objects to probe such a
precession~\cite{nor87,sck+13}.  We follow the methodology
in~\cite{sck+13} to relate the precession with observables in pulse
profile.  The precession will manifest itself in terms of a change in
the angle $\lambda$, which is defined to be the angle between
$\hat{\bf K}$ and $\hat{\bf S}$ (see Fig.~\ref{fig:pulsar_geometry}).
Purely from geometry one has, $\dot\lambda = \hat{\bf e} \cdot {\bf
  \Omega}^{\rm prec} = \pi \bar s^{jk} \hat{S}^j \hat{e}^k/
P$~\cite{sck+13}, where the unit vector $\hat{\bf e} \equiv \hat{\bf
  K} \times \hat{\bf S} / |\hat{\bf K} \times \hat{\bf S}|$ gives the
line of nodes associated with the intersection of the equatorial plane
of the pulsar and the sky plane.  Further, we introduce a simple
pulsar emission model, the cone model~\cite{lk05}, to relate the
change in $\lambda$ with the change in pulse width. Details can be
found in~\cite{sm,sck+13}.  Notice that $\hat{\bf e} \perp \hat{\bf
  S}$ by definition, therefore we are insensitive to $\bar s^{\rm TT}
= \bar s^{\rm XX} + \bar s^{\rm YY} + \bar s^{\rm ZZ}$ with solitary
pulsars.

%=====================================================================
\paragraph{Orbital dynamics of binary pulsars.\label{sec:binary}} 
%=====================================================================

In presence of LV effects, the orbital dynamics of a binary is
modified~\cite{bk06}. By using the technique of osculating elements,
Bailey and Kosteleck\'y calculated secular changes for orbital
elements~\cite{bk06,sm}. We here give an equivalent, but more compact,
vectorial form for ${\bf e} \equiv e \hat{\bf a}$ and ${\bf l} \equiv
\sqrt{1-e^2} \hat{\bf c}$~\cite{de92,sw12} after averaging over an
orbit,
%---------------------------------------------------------------------
\begin{eqnarray}
  \label{eq:edot}
  \left\langle \frac{{\rm d} \bf e}{{\rm d} t}\right\rangle &=&
  e\dot{\omega}_{\rm R} \hat{\bf b} \\
  && \hspace{-1cm} - n_b eF_e \left( F_e\sqrt{1-e^2} \bar s^{ab} \hat{\bf
      a} - F_e \frac{\bar s^{aa}- \bar s^{bb}}{2} 
    \hat{\bf b} + \bar s^{bc} \hat{\bf c} \right) \nonumber \\
  && \hspace{-1cm} + 2\delta X \, {\cal V}_{\rm O} n_b F_e \left(
    \sqrt{1-e^2} \bar s^{0a} \hat{\bf a} + \bar s^{0b} \hat{\bf b} -
    \frac{e^2}{\sqrt{1-e^2}} \bar s^{0c} \hat{\bf c}\right) \,,
  \nonumber \\
  %-----------------------------------------------------------------
  \label{eq:ldot}
  \left\langle \frac{{\rm d} \bf l}{{\rm d} t}\right\rangle &=& n_b F_e
  \left(\sqrt{1-e^2} \bar s^{bc} \hat{\bf a} - \bar s^{ac} \hat{\bf b}
    + e^2 F_e \bar s^{ab} \hat{\bf c}\right) \nonumber \\
  && + 2\delta X \, {\cal V}_{\rm O} n_b e F_e \left( \bar s^{0c} \hat{\bf
      a} - \bar s^{0a} \hat{\bf c} \right) \,,
\end{eqnarray}
%---------------------------------------------------------------------
where $\dot\omega_{\rm R}$ is the periastron advance rate, $n_b \equiv
2\pi/P_b$ is the orbital frequency, ${\cal V}_{\rm O} \equiv
\left[G(m_1+m_2)n_b\right]^{1/3}$ is the characteristic orbital
velocity, and $\delta X \equiv \frac{m_1 - m_2}{m_1 + m_2}$ denotes
the difference of the pulsar mass $m_1$ and the companion mass
$m_2$. The function $F_e \equiv \frac{1}{1+\sqrt{1-e^2}}$ only depends
on the eccentricity $e$, and $F_e \in [\frac{1}{2},1)$ for a bound
  orbit.

In pulsar timing, in principle one can construct three tests per
binary to constrain LV effects, by utilizing $\dot\omega$, $\dot e$,
and $\dot x$ ($x$ is the projected semimajor axis of pulsar
orbit)~\cite{bk06,sm}. Through a direct check, one can
show that they are insensitive to $\bar s^{\rm TT}$ as well.

%=====================================================================
\paragraph{Simulations and results.\label{sec:results}} 
%=====================================================================

%---------------------------------------------------------------------
\begin{table}
  \caption{Pulsar constraints on the coefficients of the pure-gravity
    sector of mSME~\cite{bk06}. The $K$-factor reflects the
    improvement over the combined limits from LLR and
    AI~\cite{cch+09}. Notice the probabilistic assumption made in the
    text.\label{tab:LV}}
  \begin{tabular}{p{3cm}p{4cm}r}
    \hline\hline
    SME coefficients & 68\% confidence level & $K$-factor \\
    \hline
    $\bar s^{\rm TX}$ & $(-5.2,\,5.3)\times10^{-9}$ & 118 \\
    $\bar s^{\rm TY}$ & $(-7.5,\,8.5)\times10^{-9}$ & 163 \\
    $\bar s^{\rm TZ}$ & $(-5.9,\,5.8)\times10^{-9}$ & 650 \\
    $\bar s^{\rm XY}$ & $(-3.5,\,3.6)\times10^{-11}$ & 42 \\
    $\bar s^{\rm XZ}$ & $(-2.0,\,2.0)\times10^{-11}$ & 70 \\
    $\bar s^{\rm YZ}$ & $(-3.3,\,3.3)\times10^{-11}$ & 42 \\
    $\bar s^{\rm XX}-\bar s^{\rm YY}$ & $(-9.7,\,10.1)\times10^{-11}$ &
    16 \\ 
    $\bar s^{\rm XX}+\bar s^{\rm YY}-2\bar s^{\rm ZZ}$ &
    $(-12.3,\,12.2)\times10^{-11}$ & 310 \\ 
    \hline
  \end{tabular}
\end{table}
%---------------------------------------------------------------------

Thirteen pulsars are chosen for tests, including profile observations
of two solitary MSPs, PSRs~B1937+21 and J1744$-$1134~\cite{sck+13},
that provide one test per pulsar, and timing observations of eleven
binaries, PSRs~B1913+16~\cite{wnt10}, B1534+12~\cite{sttw02},
J0737$-$3039A~\cite{ksm+06a}, B2127+11C~\cite{jcj+06},
J1738+0333~\cite{fwe+12}, J1012+5307~\cite{lwj+09},
J0348+0432~\cite{afw+13}, J1802$-$2124~\cite{fsk+10},
J0437$-$4715~\cite{vbs+08}, B1855+09, and J1909$-$3744~\cite{vbc+09},
that provide two or three tests per system.  In total we constructed
twenty-seven tests~\cite{sm}. 

Specifically, the null detection of any change in the morphology of
the pulse profiles of two solitary pulsars, after being monitored for
more than one decade~\cite{sck+13}, tightly constrains the LV spin
precession. For binary pulsars, we use the published timing solutions
of the successful phase-coherent fittings to times of arrival of pulse
signals with generic timing
models~\cite{wnt10,sttw02,ksm+06a,jcj+06,fwe+12,lwj+09,afw+13,fsk+10,vbs+08,vbc+09}. The
observations extend from several years to decades. The null detection
of any beyong-GR effects in binary pulsars constrains LV orbital
dynamics.  All these pulsar systems were studied in great detail in
their original publications, and relevant results are reviewed and
discussed in~\cite{sm}.

We made following considerations in our calculation. i)~Because
usually $\dot x$ and $\dot e$ were not reported in literature, we
conservatively estimate 68\% CL upper limits for them from
uncertainties of $e$ and $x$, as $|\dot e|^{\rm upper} = \sqrt{12}
\sigma_e /T_{\rm obs}$ and $|\dot x|^{\rm upper} = \sqrt{12} \sigma_x
/T_{\rm obs}$~\cite{sm}, where $T_{\rm obs}$ is the time span used in
deriving the timing solution, in accordance with the case of
linear-in-time evolution. The estimation is consistent with the values
reported in~\cite{sttw02} for PSR~B1534+12.  We also account for the
contribution to $\dot x$ from proper motion~\cite{kop96}.  ii)~For
consistency, the $\dot\omega$ test is possible only if component
masses are measured to a high precision independent of gravity
theories. Such mass measurements are possible only with a few
small-eccentricity binary pulsars with optical observations of
companions~\cite{lwj+09,fwe+12,afw+13}. Because these binaries have no
$\dot\omega$ measurements yet, we use the limits on time variations of
the eccentricity vector in tests with an $\dot\omega$ calculated from
GR, similar to the method proposed in~\cite{sw12}. We don't use
$\dot\omega$ in tests for pulsars whose masses were based on GR. In
contrast, $\dot e$ and $\dot x$ tests are still feasible with them
because in GR the changes in $x$ and $e$ introduced by gravitational
damping are negligible~\cite{lk05}, hence only modest knowledge on
component masses is sufficient.  iii)~One caution in directly using
Eqs.~(\ref{eq:edot}--\ref{eq:ldot}) was pointed out in~\cite{wk07}
that a large $\dot \omega$ can render the secular changes nonconstant.
These effects cannot be too large based on the fact that all binaries
were well fitted with simple timing models. The largest change in
$\omega$ is $\sim 100^\circ$ for PSR~B1913+16 in our
samples. Therefore, we consider it safe to use time-averaged values
for $\omega$-related quantities as a rough approximation at current
stage. iv)~The geometry of above pulsar systems is not fully
determined from observations. For binary pulsars, the longitude of
ascending node, $\Omega$, is generally not an observable in pulsar
timing, while for solitary pulsars, the azimuthal angle of the spin,
$\eta$, is unknown. We have to treat them as random variables
uniformly distributed in $[0^\circ,360^\circ)$. This choice makes our
  tests probabilistic tests.

\begin{figure}
  \includegraphics[width=9cm]{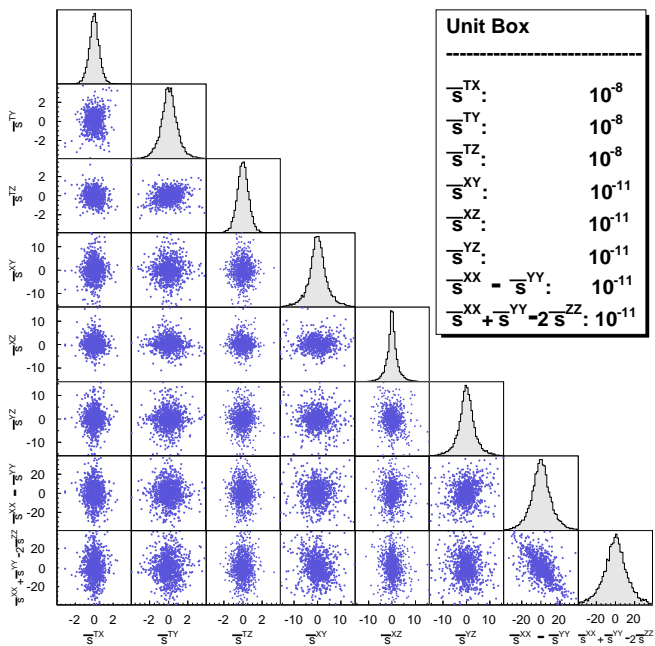}
  \caption{Correlations of LV coefficients and marginalized
    probability densities. Only $10^3$ points out of $10^4$
    simulations are plotted in each panel for
    clarity.\label{fig:corr}}
\end{figure}

We set up Monte Carlo simulations to treat unknown angles and
measurement uncertainties. First, we make lots of trials to identify
the eight most stringent tests. Linear equations for LV coefficients
are constructed from them. The equation set is solved to obtain eight
combinations of LV coefficients. Afterwards we check whether these
values are also consistent with the other nineteen tests. If all tests
are passed, values are stored. We accumulate $10^4$ entries and read
out the 68\% CLs for LV coefficients.  Results are tabulated in
Table~\ref{tab:LV}, where a comparison with that from the combination
of Lunar Laser Ranging (LLR) and atom interferometry
(AI)~\cite{cch+09} is made.  The {\it improvement factor}, $K$, is
defined to be the (inverse) ratio of the length of 68\% CLs. In
Fig.~\ref{fig:corr}, $10^3$ entries out of our results are
plotted. The absence of obvious correlations between different
coefficients benefits from multiple pulsars which make it unlikely for
a specific combination of large LV coefficients to pass {\it all}
tests.

%---------------------------------------------------------------------
%---------------------------------------------------------------------

%=====================================================================
\paragraph{Discussions.\label{sec:discussion}} 
%=====================================================================

In the parametrized post-Newtonian (PPN) formalism, there are two
parameters, $\alpha_1$ and $\alpha_2$, together with an absolute
velocity with respect to a preferred frame, ${\bf w}$, to describe
semiconservative LV effects~\cite{will93,will06}.  We notice that if
we do the replacements in the pure-gravity sector of mSME, $\bar
s^{0j} \to \alpha_1 w^j/4$ and $\bar s^{jk} \to - \alpha_2 w^j w^k$,
the dynamics of PPN is exactly recovered. We can clearly see the
relation from Lagrangians or from a comparison between
Eqs.~(\ref{eq:edot}--\ref{eq:ldot}) and Eqs.~(13--19)
in~\cite{sw12}. This new relation was also discussed by Bailey and
Kosteleck\'y in a different manner, see Eq.~(68) in~\cite{bk06}.

In scalar-tensor theories, strong fields associated with neutron stars
may develop nonperturbative effects~\cite{de93}. One would expect
similar effects in SME. Therefore our results in Table~\ref{tab:LV}
are, strictly speaking, {\it effective} constraints that include
strong-field contributions.

The tests here are of probabilistic nature because of unknown angles
$\Omega$ and $\eta$.  High-precision timing of nearby pulsars
(e.g. PSR~J0437$-$4715~\cite{vbs+08}) and observations of interstellar
scintillation (e.g. PSR~B1534+12~\cite{bplw02}) may determine
$\Omega$. When more such measurements are available, robust tests can
be performed.

All tests in this Letter are based on time derivatives of observables,
whose precisions improve as $T_{\rm obs}^{-3/2}$. Continuous
observations can quickly improve all tests.  Especially in the era of
FAST and SKA telescopes, more pulsars and high sensitivities will
further enable better tests of gravity.  We also notice that some
timing results used here were reported more than one decade ago. New
results for sure have already improved a lot. We hope observers report
$\dot e$ and $\dot x$ measurements whenever feasible, which can give
more realistic constraints for LV coefficients. It is also possible to
use dissipative effects~\cite{ybby13} once gravitational radiation is
calculated in SME.

Recently, Bailey et al.~\cite{beo13} obtained the first constraint of
the $\bar s^{\rm TT} = \bar s^{\rm XX} + \bar s^{\rm YY} +\bar s^{\rm
  ZZ}$ component from Gravity Probe B. Their result $|\bar s^{\rm TT}|
< 3.8 \times 10^{-3}$ (68\% CL) can be combined with ours to break all
degeneracy. We also note that the boost between a pulsar system and
the Solar System can be used to mix $\bar s^{\rm TT}$ with other
components; it hence provides the possibility to obtain a constraint
of $\bar s^{\rm TT}$ directly from pulsars. Currently, only a few
pulsars with optical observations of their companions have 3\,D
velocity measurements that are needed for the boost calculation. We
expect more measurements will achieve the goal in the future.

%=====================================================================
\begin{acknowledgments}
  We thank Norbert Wex for stimulating discussions and carefully
  reading the manuscript. We are grateful to Jay Tasson for
  encouragements, Joris Verbiest for discussions, and three anonymous
  referees for comments. Lijing Shao is supported by China Scholarship
  Council (CSC).
\end{acknowledgments}
%=====================================================================

%=====================================================================

\clearpage

%=====================================================================
\begin{center}
  {\Large \bf Supplemental Material}
  \\ \vspace{0.3cm} Lijing Shao (MPIfR \& PKU)
\end{center}
\setcounter{equation}{0}
\setcounter{table}{0}
%=====================================================================

%=====================================================================
\section{Rotation matrices}
%=====================================================================

The transformation matrix ${\cal R}$ from the Sun-centered
celestial-equatorial coordinate system, $(\hat{\bf X},\hat{\bf
  Y},\hat{\bf Z})$, to the binary coordinate system, $(\hat{\bf
  a},\hat{\bf b},\hat{\bf c})$, can be decomposed into five simple
parts,
%---------------------------------------------------------------------
\begin{equation}
  \left(
  \begin{array}{c}
    \hat{\bf a} \\
    \hat{\bf b} \\
    \hat{\bf c}
  \end{array}
  \right) = 
  {\cal R}
  \left(
  \begin{array}{c}
    \hat{\bf X} \\
    \hat{\bf Y} \\
    \hat{\bf Z}
  \end{array}
  \right) \,,
\end{equation}
%---------------------------------------------------------------------
with
%---------------------------------------------------------------------
\begin{equation}
  {\cal R} = {\cal R}^{(\omega)} {\cal R}^{(i)} {\cal R}^{(\Omega)}
  {\cal R}^{(\delta)} {\cal R}^{(\alpha)} \,.
\end{equation}
%---------------------------------------------------------------------
For a binary pulsar, explicit expressions of these matrices are
%---------------------------------------------------------------------
\begin{eqnarray}
  {\cal R}^{(\alpha)} &=& 
  \left(
  \begin{array}{ccc}
    -\sin\alpha & \cos\alpha & 0 \\
    -\cos\alpha & -\sin\alpha & 0 \\
    0 & 0 & 1
  \end{array}
  \right) \,,\\
  %-----------------------------------------------------------
  {\cal R}^{(\delta)} &=& 
    \left(
  \begin{array}{ccc}
    1 & 0 & 0 \\
    0 & \sin\delta & \cos\delta \\
    0 & -\cos\delta & \sin\delta
  \end{array}
  \right) \,,\\
  %-----------------------------------------------------------
  {\cal R}^{(\Omega)} &=&
    \left(
  \begin{array}{ccc}
    \cos\Omega & \sin\Omega & 0 \\
    -\sin\Omega & \cos\Omega & 0 \\
    0 & 0 & 1
  \end{array}
  \right) \,,\\
  %-----------------------------------------------------------
  {\cal R}^{(i)} &=& 
    \left(
  \begin{array}{ccc}
    1 & 0 & 0 \\
    0 & \cos i & \sin i \\
    0 & -\sin i & \cos i
  \end{array}
  \right) \,,\\
  %-----------------------------------------------------------
  {\cal R}^{(\omega)} &=& 
    \left(
  \begin{array}{ccc}
    \cos\omega & \sin\omega & 0 \\
   -\sin\omega & \cos\omega & 0 \\
    0 & 0 & 1
  \end{array}
  \right) \,,
\end{eqnarray}
%---------------------------------------------------------------------
where $\alpha$ (right ascension), $\delta$ (declination), $\Omega$
(longitude of ascending node), $i$ (orbital inclination), $\omega$
(longitude of periastron) are defined in Figs.~1--2 in the main text.

For a solitary pulsar, only a rotation from $(\hat{\bf X},\hat{\bf
  Y},\hat{\bf Z})$ to $(\hat{\bf I},\hat{\bf J},\hat{\bf K})$ is
needed, hence one can set ${\cal R}^{(\Omega)}$, ${\cal R}^{(i)}$, and
${\cal R}^{(\omega)}$ to unit matrices.

%=====================================================================
\section{Pulsar emission model and LV spin precession}
%=====================================================================

As in~\cite{sck+13}, we adopt the cone model for pulsar
emission~\cite{lk05}. Other choices only affect the results
marginally. In the cone model, from the geometry, one has,
%---------------------------------------------------------------------
\begin{equation}\label{sm:eq:lambda1}
  \sin^2 \left(\frac{W}{4}\right) = \frac{\sin^2(\rho/2)
    -\sin^2(\beta/2)}{\sin(\alpha+\beta) \sin\alpha} \,,
\end{equation}
%---------------------------------------------------------------------
where $W$ is the width of the pulse, $\alpha$ is the magnetic
inclination angle, $\beta \equiv 180^\circ - \lambda - \alpha$ is the
impact angle, and $\rho$ is the semi-angle of the opening radiating
region.  Adopting a plausible assumption that the radiation property
has no change during the observational span, i.e. ${\rm d}\alpha /
{\rm d}t = {\rm d}\rho / {\rm d}t = 0$, one has~\cite{sck+13},
%---------------------------------------------------------------------
\begin{equation}\label{sm:eq:lambda2}
  \frac{{\rm d} \lambda}{{\rm d} t} = \frac{1}{2} \frac{\sin
    (W/2)}{\cot\lambda \cos(W/2)+\cot\alpha} \frac{{\rm d} W}{{\rm d}
    t} \,.
\end{equation}
%---------------------------------------------------------------------
Combining it with the LV spin precession rate, $\Omega_k^{\rm prec} =
\pi \bar s^{jk} \hat S^j / P$, and the geometric relation,
$\dot\lambda = \hat{\bf e}\cdot{\bf \Omega}^{\rm prec}$ (see the main
text), one can relate the LV coefficients with the time derivative of
the pulse width,
%---------------------------------------------------------------------
\begin{equation}\label{sm:eq:dwdt}
  \frac{{\rm d} W}{{\rm d} t} = \frac{2\pi}{P}\frac{\cot\lambda
    \cos(W/2)+\cot\alpha}{\sin(W/2)} \bar s^{jk} \hat{S}^j \hat e^k
  \,.
\end{equation}
%---------------------------------------------------------------------

%=====================================================================
\section{Orbital dynamics of a binary in presence of LV effects}
%=====================================================================

We collect from~\cite{bk06} with slightly different notations the
orbital-averaged secular changes of $a$, $e$, $\omega$, and $x$,
%---------------------------------------------------------------------
\begin{eqnarray}
  \left\langle \frac{{\rm d} a}{{\rm d} t}\right\rangle &=&
  0 \label{sm:eq:adot} \,, \\
  %------------------------------------------------------------
  \left\langle \frac{{\rm d} e}{{\rm d} t}\right\rangle
  &=& n_b F_e \sqrt{1-e^2} \left( -eF_e \bar s^{ab} + 2\delta X \,
    {\cal V}_{\rm O} \bar s^{0a}\right) \label{sm:eq:edot} \,, \\
  %------------------------------------------------------------
  \left\langle \frac{{\rm d} \omega}{{\rm d} t}\right\rangle &=&
  \frac{3n_b{\cal V}_{\rm O}^2}{1-e^2} - \frac{n_b F_e \cot
    i}{\sqrt{1-e^2}} \times \\ 
  && \hspace{-1cm} \left(\sin\omega \, \bar s^{ac} +
    \sqrt{1-e^2}\cos\omega \,\bar 
    s^{bc}  + 2\delta X \, e{\cal V}_{\rm O} \cos\omega \, \bar
    s^{0c} \right) \nonumber \\
  && \hspace{-1cm} + n_b F_e\left( F_e \frac{\bar
    s^{aa}-\bar s^{bb}}{2} + \frac{2}{e}\delta X \,  {\cal
      V}_{\rm O}\bar s^{0b} \right) \label{sm:eq:omdot} \,, \nonumber \\
  %------------------------------------------------------------
  \label{sm:eq:xdot} \left\langle \frac{{\rm d} x}{{\rm d}
    t}\right\rangle &=& \frac{1-\delta X}{2}\frac{F_e{\cal
    V}_{\rm O}\cos i}{\sqrt{1-e^2}}  \times  
    \\ &&  \hspace{-1cm} \left(
     \cos\omega \, \bar s^{ac}
    - \sqrt{1-e^2}\sin\omega \, \bar s^{bc}  -2\delta X \,e {\cal
      V}_{\rm O}\sin\omega \, \bar s^{0c} \right) 
    \,,\nonumber 
\end{eqnarray}
%---------------------------------------------------------------------
where $a$ is the semimajor axis of the {\it orbit} and $x$ is the
projected semimajor axis of the {\it pulsar}. Other notations are
given in the main text. These equations are equivalent to Eqs.~(2--3)
in the main text.

In the limit of small eccentricity ($e \ll 1$), above equations
reduce to
%---------------------------------------------------------------------
\begin{eqnarray}
  \left\langle \frac{{\rm d} a}{{\rm d} t}\right\rangle &=& 0
  \,, \label{sm:eq:ell1adot} \\
  %-------------------------------------------------------------------
  \left\langle \frac{{\rm d} e}{{\rm d} t}\right\rangle &\simeq& n_b
  \delta X \, {\cal V}_{\rm O} \bar s^{0a} \,,\label{sm:eq:ell1edot} \\
  %-------------------------------------------------------------------
  \left\langle \frac{{\rm d} \omega}{{\rm d} t}\right\rangle &\simeq&
  3n_b{\cal V}_{\rm O}^2 + \frac{n_b}{e} \delta X \, {\cal V}_{\rm O} \bar
  s^{0b} \,, \label{sm:eq:ell1omdot}\\ 
  %-------------------------------------------------------------------
  \label{sm:eq:ell1xdot} \left\langle \frac{{\rm d} x}{{\rm d}
    t}\right\rangle &\simeq& 
  \frac{1-\delta X}{4} {\cal V}_{\rm O}\cos i \left( \bar s^{ac}
  \cos\omega - \bar s^{bc} \sin\omega \right) \,. 
\end{eqnarray}
%---------------------------------------------------------------------
Time derivatives of two Laplace-Lagrange parameters are easily
obtained from Eqs.~(\ref{sm:eq:ell1adot}--\ref{sm:eq:ell1xdot}),
%---------------------------------------------------------------------
\begin{eqnarray}
  \hspace{-1cm} \left\langle \frac{{\rm d} \eta}{{\rm d}
    t}\right\rangle &\simeq& n_b \delta X {\cal V}_{\rm O} \left( \bar
  s^{0a}\sin\omega + \bar s^{0b}\cos\omega\right) + 3e n_b {\cal
    V}_{\rm O}^2 \cos\omega\,, \\
  %------------------------------------------------------------
  \hspace{-1cm} \left\langle \frac{{\rm d} \kappa}{{\rm d}
    t}\right\rangle &\simeq& n_b \delta X {\cal V}_{\rm O} \left( \bar
  s^{0a}\cos\omega - \bar s^{0b}\sin\omega\right) -3en_b {\cal V}_{\rm
    O}^2 \sin\omega\,,
\end{eqnarray}
%---------------------------------------------------------------------
where $\eta \equiv e\sin\omega$ and $\kappa \equiv e\cos\omega$.

%=====================================================================
\section{A brief overview of pulsar systems used in LV tests}
%=====================================================================

%---------------------------------------------------------------------
\begin{figure*}
  \includegraphics[width=15cm]{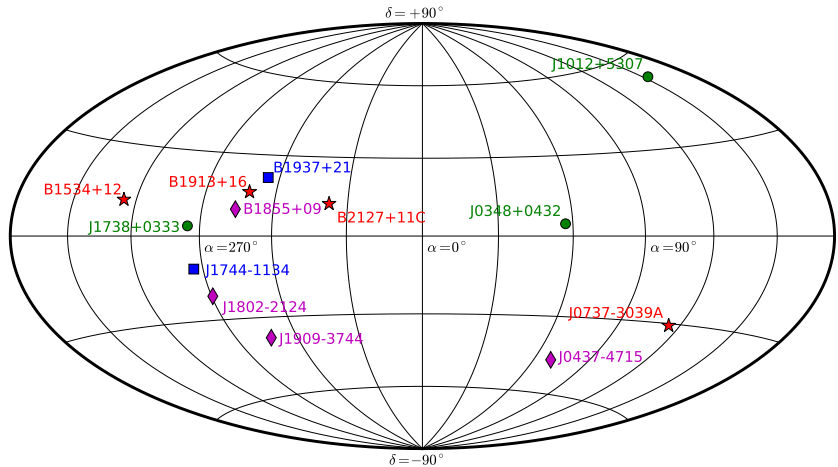}
  \caption{Distributions of pulsar systems in the sky used in tests of
    the pure-gravity sector of mSME. Blue squares, green circles,
    magenta diamonds, and red stars denote solitary pulsars,
    small-eccentricity binary pulsars with theory-independent mass
    measurements, small-eccentricity binary pulsars without
    theory-independent mass measurements, and eccentric binary pulsars,
    respectively.\label{sm:fig:pulsar_sky}}
\end{figure*}
%---------------------------------------------------------------------

In testing the pure-gravity sector of mSME, 13 pulsar systems are
used. The binaries are selected mainly based on their orbital
compactness and timing precision. From equations above, it is easy to
see that binary pulsars with short orbital periods are perferred to
perform LV tests. The distribution of our pulsars in the sky is
plotted in Fig.~\ref{sm:fig:pulsar_sky}. The wide coverage with multiple
pulsar systems, in terms of positional direction and orbital geometry,
helps to disentangle the degeneracy of LV coefficients~\cite{bk06}.

Relevant parameters of our pulsar systems for LV tests are tabulated
in Tables~\ref{sm:tab:solitary}--\ref{sm:tab:psr:ecc}.  We have grouped them
into solitary pulsars (PSRs~B1937+21 and J1744$-$1134~\cite{sck+13};
Table~\ref{sm:tab:solitary}), small-eccentricity binary pulsars with
theory-independent mass measurements (PSRs~J1012+5307~\cite{lwj+09},
J1738+0333~\cite{fwe+12}, and J0348+0432~\cite{afw+13};
Table~\ref{sm:tab:psr:smalle}), small-eccentricity binary pulsars without
theory-independent mass measurements (PSRs J1802$-$2124~\cite{fsk+10},
J0437$-$4715~\cite{vbs+08}, B1855+09, and J1909$-$3744~\cite{vbc+09};
Table~\ref{sm:tab:psr:smalle2}), and eccentric binary pulsars
(PSRs~B1913+16~\cite{wnt10}, B1534+12~\cite{sttw02},
B2127+11C~\cite{jcj+06}, and J0737$-$3039A~\cite{ksm+06a};
Table~\ref{sm:tab:psr:ecc}).  Brief reviews of each pulsar system are
given below, and interested readers are pointed to original
publications and references therein for details.

By using the phrase ``{\it pulsars with theory-independent mass
  measurements}'', we refer to those binary pulsars that have mass
measurements only based on the weak-field gravity theory, whose
validity has already been well-tested in the Solar
system~\cite{will06}. In our samples, three small-eccentricity binary
pulsars have such mass measurements~\cite{lwj+09,fwe+12,afw+13}. The
optical observations of their white dwarf (WD) companions give the
masses of WDs from well-established WD models. Combining the optical
and radio observations, one can obtain the mass ratio, $q\equiv
m_1/m_2$, from the amplitudes of their projected orbital velocities
along the line of sight.  In contrast, ``{\it pulsars without
  theory-independent mass measurements}'' refer to binary pulsars that
have masses derived from post-Newtonian effects of GR. If these
effects with strongly self-gravitating bodies were modified in
alternative gravity theories, the GR masses cannot be trusted with
high confidence in precision tests of alternative gravity
theories. For the reasons stated in the main text, we can use
component masses derived from GR as an approximation in $\dot e$ and
$\dot x$ tests.

In our LV tests, solitary pulsars provide one test per system with
Eq.~(\ref{sm:eq:dwdt}), and binary pulsars with theory-independent mass
measurements provide three tests per system with
Eq.~(\ref{sm:eq:ell1xdot}) and the time variations of the eccentricity
vector\footnote{With the new relation between the pure-gravity sector
  of mSME and the PPN framework in the main text, it is easy to check
  that for small-eccentricity binary pulsars, the time-spatial
  components of $\bar s^{\mu\nu}$ introduce ``{\it orbital
    polarization effects}'', in a similar way to that introduced by
  the PPN parameter $\alpha_1$~\cite{sw12}. This can also be easily
  checked from Eqs.~(\ref{sm:eq:ell1edot}) and (\ref{sm:eq:ell1omdot}).},
while bianry pulsars without theory-independent mass measurements
provide two tests per system with Eqs.~(\ref{sm:eq:edot}) and
(\ref{sm:eq:xdot}).  Therefore, in total, we have performed 27 tests on 8
linear combinations of LV coefficients.

%=====================================================================
\subsection{Solitary pulsars}
%=====================================================================

%=====================================================================
\begin{table*}
  \caption{Relevant quantities of PSRs~B1937+21 and J1744$-$1134 for
    the test. Most quantities are from pulsar timing, while the
    orientation and radiation quantities ($\alpha$ and $\zeta$) were
    obtained from the model fitting to radio and $\gamma$-ray
    lightcurves (see~\cite{sck+13} and references therein). For
    PSR~B1937+21, quantities for the main-pulse (left) and the
    interpulse (right) are both tabulated.  Parenthesized numbers
    represent the $1$-$\sigma$ uncertainty in the last digits
    quoted. \label{sm:tab:solitary}}
\begin{center}
  \begin{tabular}{p{8cm}p{4.5cm}p{4.5cm}}
    \hline\hline
    Pulsar & PSR~B1937+21 & PSR~J1744$-$1134 \\
    \hline
    Discovery (year) & 1982 &  1997 \\
    Right Ascension, $\alpha$ (J2000) &
    $19^{\rm h}39^{\rm m}38^{\rm s}\!.561297(2)$ &
    $17^{\rm h}44^{\rm m}29^{\rm s}\!.403209(4)$ \\ 
    Declination, $\delta$ (J2000) &
    $+21^\circ34'59''\!.12950(4)$ &
    $-11^\circ34'54''\!.6606(2)$ \\ 
    Spin period, $P$ (ms) & 1.55780653910(3) & 4.074545940854022(8) \\
    Proper motion in $\alpha$, $\mu_\alpha$ (mas\,yr$^{-1}$) &
    $0.072(1)$ & 18.804(8) \\ 
    Proper motion in $\delta$, $\mu_\delta$ (mas\,yr$^{-1}$) &
    $-0.415(2)$ & $-9.40(3)$ \\
    Magnetic inclination, $\alpha$ (deg) & $75^{+8}_{-6}$ ~~/~~
    $105^{+6}_{-8}$ & $51^{+16}_{-19}$ \\ 
    Observer angle, $\zeta \equiv 180^\circ - \lambda$ (deg) & $80(3)$
    & $85^{+3}_{-12}$ \\  
     Time span of data (MJD) & 50693--55725 & 50460--55962\\ 
    Pulse width at 50\% intensity, $W_{50}$ (deg) & 8.281(9) ~~/~~
    10.245(17) & 12.53(3) \\ 
    Time derivative of $W_{50}$, ${\rm d}{W}_{50}/{\rm d} t$
    ($10^{-3}\,\mbox{deg}\,{\rm yr}^{-1}$) & $-3.2(34)$ ~~/~~ 3.5(66) &
    $1.3(72)$ \\ 
    \hline
  \end{tabular}
\end{center}
\end{table*}
%=====================================================================

%=====================================================================
\subsubsection{PSR B1937+21}
\label{sm:sec:B1937}
%=====================================================================

PSR~B1937+21 (a.k.a. PSR J1939+2134) is the first discovered MSP, with
a spin period of 1.56\,ms.  It is a bright pulsar in the radio band
and has a stable rotation.  Therefore, PSR~B1937+21 was chosen as an
important target in the pulsar timing array (PTA) projects, and has
been observed continuously and frequently since its discovery in
1982. Although with substantial low-frequency red noises, the timing
residual has reached a level $\lesssim 0.4$\,$\mu$s~\cite{vbc+09}.
Recently, Shao et al.~\cite{sck+13} analyzed $\sim15$ years of
archival data, taken at the 100-m Effelsberg radio telescope, to look
into the stability of the pulse profile of this pulsar. The profile
consists of a main-pulse and an interpulse. They are both very stable
with their time derivatives of width constrained to be
$(-3.2\pm3.4)\times10^{-3}$\,deg\,yr$^{-1}$ and
$(3.5\pm6.6)\times10^{-3}$\,deg\,yr$^{-1}$ respectively~\cite{sck+13}.
The null change in pulse width indicates a null LV spin precession.

%=====================================================================
\subsubsection{PSR J1744$-$1134}
\label{sm:sec:J1744}
%=====================================================================

PSR~J1744$-$1134 was discovered in 1997 through the Parkes 436\,MHz
survey of the southern sky. The pulsar has a spin period of 4.07\,ms,
and later was chosen as a target in PTA projects, being observed
frequently as well.  The timing residual reaches sub-$\mu$s
in~\cite{vbc+09}.  This pulsar has a clear main-pulse, and the pulse
profile is very stable against time, whose time derivative of the
pulse width was constrained to be
$(1.3\pm7.2)\times10^{-3}$\,deg\,yr$^{-1}$ from 15 years of Effelsberg
data~\cite{sck+13}.

%=====================================================================
\subsection{Small-eccentricity binary pulsars with theory-independent mass
  measurements} 
%=====================================================================

%---------------------------------------------------------------------
\begin{table*}
\caption{Relevant quantities of PSRs~J1012+5307 \cite{lwj+09},
  J1738+0333 \cite{fwe+12}, and J0348+0432~\cite{afw+13} for the test,
  from radio-timing and optical observations. Parenthesized numbers
  represent the $1$-$\sigma$ uncertainty in the last digits
  quoted. The listed Laplace-Lagrange parameter, $\eta$, is the
  intrinsic value, after subtraction of the contribution from the
  Shapiro delay. There is an ambiguity between $i$ and $180^\circ-i$;
  only the value $i < 90^\circ$ is tabulated. \label{sm:tab:psr:smalle}}
  \begin{tabular}{p{6.5cm}p{3.5cm}p{3.5cm}p{3.5cm}}
    \hline\hline
    Pulsar & PSR~J1012+5307 & PSR~J1738+0333 & PSR~J0348+0432 \\
    \hline
    \multicolumn{4}{l}{\tt Observed Quantities} \\
    \hline
    Observational span, $T_{\rm obs}$ (year) & $\sim15$~\cite{lwj+09} &
    $\sim10$~\cite{fwe+12} & $\sim4$~\cite{afw+13} \\
    Right ascension, $\alpha$ (J2000) &
      ${\rm 10^h12^m33^s \hspace{-1.2mm}.  4341010(99)}$ &
      ${\rm 17^h38^m53^s \hspace{-1.2mm}. 9658386(7)}$ &
      ${\rm 03^h48^m43^s \hspace{-1.2mm}. 639000(4)}$ \\
    Declination, $\delta$ (J2000) &
      $53^\circ07^\prime02^{\prime\prime} \hspace{-1.2mm} .  60070(13)$ & 
      $03^\circ33^\prime10^{\prime\prime} \hspace{-1.2mm} .  86667(3)$&
      $04^\circ32^\prime11^{\prime\prime} \hspace{-1.2mm} .  4580(2)$ \\
    Proper motion in $\alpha$, $\mu_\alpha~(\textrm{mas\,yr}^{-1})$ & 
    2.562(14) & 7.037(5) & 4.04(16) \\
    Proper motion in $\delta$, $\mu_\delta~(\textrm{mas\,yr}^{-1})$ & 
    $-$25.61(2) & 5.073(12) & 3.5(6) \\
    Spin period, $P$ (ms) & 5.255749014115410(15) &
    5.850095859775683(5) & 39.1226569017806(5) \\
    Orbital period, $P_{\rm b}$ (day) & 0.60467271355(3) & 0.3547907398724(13)
    & 0.102424062722(7) \\
    Projected semimajor axis, $x$ (lt-s) & 0.5818172(2) &
    0.343429130(17) & 0.14097938(7) \\
    $\eta \equiv e \sin\omega~(10^{-7})$ & $-1.4\pm3.4$ &  
    $-1.4\pm1.1$ & $19\pm10$ \\
    $\kappa \equiv e \cos\omega~(10^{-7})$ & $0.6\pm3.1$ & $3.1\pm1.1$
    & $14\pm10$ \\
    Time derivative of $x$, $\dot{x}~(10^{-15}~\textrm{s\,s}^{-1})$ & 2.3(8) & 
    0.7(5) & $\cdots$ \\
    Mass ratio, $q \equiv m_1/m_2$ & 10.5(5) & 8.1(2) & 11.70(13) \\
    Companion mass, $m_2~({\rm M}_\odot)$ & 0.16(2) &
    $0.181^{+0.008}_{-0.007}$ & 0.172(3) \\
    Pulsar mass, $m_1~(\textrm{M}_\odot)$ & 1.64(22) &
    $1.46^{+0.06}_{-0.05}$ & 2.01(4) \\
    $\delta X \equiv  (q-1)/(q+1)$ & 0.826(8) & 0.780(5) &
    0.843(2) \\
    \hline
    \multicolumn{4}{l}{\tt Estimated Quantities} \\
    \hline
    Upper limit of $|\dot x|~(10^{-15}~\textrm{s\,s}^{-1})$ & $\cdots$
    & $\cdots$ & 1.9 \\
    Upper limit of $|\dot\eta|$ ($10^{-14}\,{\rm s}^{-1}$) & 0.25 & 0.12
    & 2.7 \\
    Upper limit of $|\dot\kappa|$ ($10^{-14}\,{\rm s}^{-1}$) & 0.23 &
    0.12 & 2.7 \\
    \hline
    \multicolumn{4}{l}{\tt Derived Quantities Based on GR} \\
    \hline
    Orbital inclination, $i$ (deg) &  52(4) & 32.6(10) & 40.2(6) \\
    Advance of periastron, $\dot{\omega}\,({\rm
      deg\,yr}^{-1})$ & 0.69(6) & 1.57(5) & 14.9(2) \\
    Characteristic velocity, $\mathcal{V}_{\rm O}\,({\rm
      km\,s}^{-1})$ & 308(13) & 355(5) & 590(4) \\
    \hline
  \end{tabular}
\end{table*}
%---------------------------------------------------------------------

%=====================================================================
\subsubsection{PSR J1012+5307}
\label{sm:sec:J1012}
%=====================================================================

PSR~J1012+5307 is a small-eccentricity neutron star (NS) white dwarf
(WD) binary system, with an orbital period of $14.5$\,hours. The
pulsar was discovered in 1993 with the 76-m Lovell radio telescope at
Jodrell Bank. Later optical observations revealed its companion being
a helium WD, and determined the mass ratio $q \equiv m_1/m_2 = 10.5
\pm 0.5$, and the WD mass $m_2 = 0.16 \pm 0.02\,\textrm{M}_\odot$. The
3\,D spatial velocity of this binary was also measured.  Lazaridis et
al.~\cite{lwj+09} presented the most updated timing solution by using
15 years of observations from the European PTA (EPTA) network, from
which they obtained stringent limits on the gravitational dipole
radiation and the time variation of the gravitational constant.  Being
a well-timed relativistic binary, PSR~J1012+5307 is useful to
constrain local Lorentz invariance violation of gravity. It was used
to constrain the PPN parameters, $\alpha_1$ and $\alpha_2$, in tests
proposed in~\cite{sw12}.

%=====================================================================
\subsubsection{PSR J1738+0333}
\label{sm:sec:J1738}
%=====================================================================

PSR~J1738+0333 was discovered in 2001 in the high Galactic latitude
survey with the 64-m Parkes telescope, and later regularly timed with
the 305-m Arecibo telescope. It is one of MSPs known to be orbited by
a WD companion bright enough for high-resolution
spectroscopy. Accurate binary parameters and the 3\,D spatial motion
for the binary can be found in~\cite{fwe+12} and references
therein. This small-eccentricity NS-WD binary has an orbital period of
$8.5$\,hours, which, together with other well measured quantities,
makes it a superb astrophysical laboratory to test gravity theories
\cite{fwe+12}.  PSR~J1738+0333 was used to constrain the PPN
parameters, $\alpha_1$ and $\alpha_2$~\cite{sw12}. The constraint on
$\alpha_1$ from this pulsar is the best limit up to now.

%=====================================================================
\subsubsection{PSR J0348+0432}
\label{sm:sec:J0348}
%=====================================================================

PSR~J0348+0432 was discovered in a 350\,MHz drift-scan survey using
the Green Bank telescope.  It has a spin period of 39\,ms and an orbit
period of 2.5\,hours.  The companion is a low-mass WD, whose spectra
were later phase-resolved by optical observations from the Very Large
Telescope~\cite{afw+13}. The mass of WD was derived based on the
comparison of optical observations with well-tested theoretical models
of WDs, which gives $m_2 = 0.172\pm0.003 \,M_\odot$. The mass ratio
$q$ was derived from the ratio of the WD radial velocity (from
phase-resolved optical observations) to the pulsar radial velocity
(from radio-timing observations). The mass ratio, along with the WD
mass, leads to $m_1 = 2.01\pm0.04\,M_\odot$, that is the highest
well-measured NS mass~\cite{afw+13}.  PSR~J0348+0432 is a superb
system for studies of equation of state of superdense nuclear matter,
binary evolution, and tests of gravity theories. In~\cite{afw+13}, the
agreement of its measured $\dot P_b$ with that from GR imposes
stringent constraints on the parameter space of scalar-tensor
theories. For studies here, it is worthy to mention that the
periastron advance rate calculated in GR is $\dot\omega \simeq
15\,{\rm deg\,yr}^{-1}$ for PSR~J0348+0432, which would have resulted
in a rotation of the periastron for $\sim55^\circ$ in $T_{\rm
  obs}\simeq3.7$\,years.

%=====================================================================
\subsection{Small-eccentricity binary pulsars without theory-independent
  mass measurements}
%=====================================================================

%---------------------------------------------------------------------
\begin{table*}
  \caption{Relevant parameters in LV tests for
    PSRs~J1802$-$2124~\cite{fsk+10}, J0437$-$4715~\cite{vbs+08},
    B1855+09, and J1909$-$3744~\cite{vbc+09}. Parenthesized numbers
    represent the uncertainty in the last quoted digits. There is an
    ambiguity between $i$ and $180^\circ-i$; only the value $i <
    90^\circ$ is tabulated.\label{sm:tab:psr:smalle2}}
\begin{tabular}{lllll}
    \hline\hline
    Pulsar & PSR~J1802$-$2124 & PSR~J0437$-$4715 & PSR~B1855+09 &
    PSR~J1909$-$3744 \\
    \hline
   \multicolumn{5}{l}{\tt Observed Quantities} \\
   \hline
   Observational span, $T_{\rm obs}$\,(year) & $\sim6$~\cite{fsk+10} &
   $\sim10$~\cite{vbs+08} & $\sim22$~\cite{vbc+09} & $\sim5$~\cite{vbc+09} \\
   Right ascension, $\alpha$ (J2000) &
   ${\rm 18^h02^m05^s \hspace{-1.2mm}. 335576(5)}$ &
   ${\rm 04^h37^m15^s \hspace{-1.2mm}. 814764(3)}$ &
   ${\rm 18^h57^m36^s \hspace{-1.2mm}. 392909(7)}$ &
   ${\rm 19^h09^m47^s \hspace{-1.2mm}. 4366120(4)}$ \\
   Declination, $\delta$ (J2000) &
   $-21^\circ24^\prime03^{\prime\prime} \hspace{-1.2mm} .  649(2)$ &
   $-47^\circ15^\prime08^{\prime\prime} \hspace{-1.2mm} .  62417(3)$ &
   $09^\circ43^\prime17^{\prime\prime} \hspace{-1.2mm} .  2754(2)$ &
   $-37^\circ44^\prime14^{\prime\prime} \hspace{-1.2mm} .  38013(2)$ \\    
   Proper motion in $\alpha$, $\mu_\alpha$ (mas\,yr$^{-1}$) &
   $-0.85$(5) & 121.45(1) & $-2.64(2)$ & $-9.510(4)$\\
   Proper motion in $\delta$, $\mu_\delta$ (mas\,yr$^{-1}$) & 
   $<2.4$ & $-71.46(1)$ & $-5.46(2)$ & $-35.859(10)$ \\
   Spin period, $P$ (ms) & 12.6475935865227(3) & 5.7574519243621(1)
   & 5.36210054173545(3) & 2.9471080681076399(5)\\ 
   Orbital period, $P_{\rm b}$ (day) & 0.698889243381(5) &
   5.741046(2) & 12.32719(2) & 1.533449474590(3) \\ 
   Projected semimajor axis, $x$ (lt-s) & 3.7188533(5) & 3.36669708(14)
   & 9.230780(3) & 1.89799106(4) \\
   Eccentricity, $e$ ($10^{-5}$)  &  0.248(5) & 1.9180(7) & 2.170(3) &
   0.013(1)\\
   Longitude of periastron, $\omega$ (deg) & $20(2)$ & 1.22(5) &
   276.5(1) & 182(9) \\
   Epoch of periastron, $T_0$ (MJD) & 53452.673(4) & 52009.8524(8) &
   50476.095(4) & 53630.723214894(2) \\
   $\eta \equiv e\sin\omega$ ($10^{-7}$) & 8.6(9) & $\cdots$ &
   $\cdots$ & $-0.04(20)$ \\
   $\kappa \equiv e\cos\omega$ ($10^{-7}$) & 23.2(4) & $\cdots$ &
   $\cdots$ & $-1.3(1)$ \\
   Shapiro delay parameter, $s$ & 0.984(2) & 0.675(3) & 0.9990(4) &
   0.9980(1) \\
   Shapiro delay parameter, $r$ ($\mu$s) & 3.8(2) & 1.25(9) & 1.33(7)
   & 1.04(1) \\
   Longitude of ascending node, $\Omega$ (deg) & $\cdots$ & 208(7) & $\cdots$
   & $\cdots$ \\ 
   \hline
   \multicolumn{5}{l}{\tt Estimated Quantities} \\
   \hline
   Upper limit of $|\dot e|$ ($10^{-16}$\,s$^{-1}$) & 9.1 & 0.77 & 1.5
   & 2.2 \\
   Upper limit of $|\dot x|$ ($10^{-15}$\,${\rm s\,s}^{-1}$) & 9.1 & 1.5
    & 15 & 0.88 \\
   \hline
   \multicolumn{5}{l}{\tt Derived Quantities Based on GR} \\
   \hline
   Pulsar mass, $m_1$ (M$_\odot$) & 1.24(11) & 1.76(20) & 1.6(2) & 1.53(2) \\
   Companion mass, $m_2$ (M$_\odot$) & 0.78(4) & 0.254(18) & 0.27(2) &
   0.212(2) \\ 
   $\delta X \equiv (m_1 - m_2) / (m_1 + m_2)$ & 0.23(5) & 0.75(3) & 0.71(1) &
   0.757(1) \\
   Inclination, $i$ (deg) & 79.9(6) & 42.4(2) & 87.5(5) & 86.4(1) \\
   Advance of periastron, $\dot\omega$\,(deg\,yr$^{-1}$) & 0.58(2) & 0.016(8) &
   0.0046(3) & 0.141(1) \\
   Characteristic velocity, ${\cal V}_{\rm O}$ (km\,s$^{-1}$) & 303(6)
   & 150(5) & 114(4) & 222(1)\\
   \hline
\end{tabular}
\end{table*}
%---------------------------------------------------------------------

%=====================================================================
\subsubsection{PSR J1802$-$2124}
%=====================================================================

PSR J1802$-$2124 was discovered in the Parkes Multibeam Pulsar Survey
in 2002.  It is a 12.6\,ms pulsar in a 16.8-hour orbit with a
relatively massive WD companion.  It is a useful example of the
intermediate-mass class of binary pulsar systems, that provide
interesting clues to binary evolution scenarios. The relatively large
companion mass and its fortunately large orbital inclination angle
produce a detectable Shapiro delay.  It was measured through
observations conducted at Parkes telescope, Green Bank telescope, and
Nan\c{c}ay telescope~\cite{fsk+10}. The {\it shape} and {\it range} of
Shapiro delay help to pin down the component masses in GR to be $m_1 =
1.24 \pm 0.11\,{\rm M}_\odot$ and $m_2 = 0.78 \pm 0.04\,{\rm
  M}_\odot$. See Table~\ref{sm:tab:psr:smalle2} for the timing solution
of PSR~J1802$-$2124~\cite{fsk+10}. It is worthy mentioning that the
precision for the orbital eccentricity has reached a level at $
\sigma_e \sim {\cal O}(10^{-8})$.

%=====================================================================
\subsubsection{PSR J0437$-$4715}
%=====================================================================

PSR J0437$-$4715 is the brightest and nearest MSP, discovered at a
Parkes survey at 430\,MHz in 1993. It is also one of the most
well-timed pulsars with a timing residual $\sim200$\,ns~\cite{vbs+08}.
Shapiro delay was detected in PSR~J0437$-$4715. Moverover, because of
its proximity, the secular change in the inclination angle of the
orbit, due to its proper motion that gradually alters our line of
sight to the orbital plane, was observed. It allows to break the
ambiguity between $i$ and $180^\circ -i$, and also determine the
longitude of ascending node, $\Omega$~\cite{kop96}. Therefore, the
3\,D orbital geometry is fully determined for PSR~J0437$-$4715. The
most updated timing solution is presented in~\cite{vbs+08} based on 10
years of Parkes high-precision observations (see
Table~\ref{sm:tab:psr:smalle2}). The very precison of this binary allowed
to detect errors in the old Solar system ephemerides and place a
stringent limit on the time variation of the gravitational
constant~\cite{vbs+08}.  For relevance here, the precision for the
orbital eccentricity has reached a level at $ \sigma_e \sim {\cal
  O}(10^{-9})$.  PSR~J0437$-$4715 is also one of the most important
target in the PTA projects aiming at a detection of gravitational
waves~\cite{vbc+09}.

%=====================================================================
\subsubsection{PSR B1855+09}
%=====================================================================

PSR~B1855+09 (a.k.a. PSR~J1857+0943) was detected by Arecibo telescope
in 1986. It is a binary pulsar with a 5.4\,ms spin period in a nearly
circular 12.3-day orbit with a helium WD companion. The WD companion
was later detected optically by Keck and Hubble Space Telescope
observations.  Shapiro delay in this system was detected in early
1990's that revealed its nearly edge-on orbit ($i \sim 88^\circ$). It
is the first detected Shapiro delay in pulsar binaries. Although LV
tests in general require relativistic orbits, PSR~B1855+09, with a
relatively long orbital period, $P_{\rm b}\sim12$\,days, still
provides some sensitivities due to its high-precision timing and a
long timing baseline $T_{\rm obs}\gtrsim 20$\,years. The updated
timing solution~\cite{vbc+09} is presented in
Table~\ref{sm:tab:psr:smalle2}.

%=====================================================================
\subsubsection{PSR J1909$-$3744}
%=====================================================================

PSR~J1909$-$3744 was discovered during the Swinburne High Latitude
Pulsar Survey using the Parkes telescope. It is a bianry MSP with a
very small duty cycle in its pulse profile, that permits
high-precision radio timing. The highly inclined orbit
($i\sim86^\circ$) allowed the measurement of the relativistic Shapiro
delay in PSR~J1909$-$3744. A timing solution from 5 years of Parkes
observations is presented in Table~\ref{sm:tab:psr:smalle2}, where the
timing residual has reached a level less than
200\,ns~\cite{vbc+09}. It is among the best timed pulsars that are
being used to detect gravitational waves in PTAs.

%=====================================================================
\subsection{Eccentric binary pulsars}
%=====================================================================

%=====================================================================
\subsubsection{PSR B1913+16}
\label{sm:sec:B1913}
%=====================================================================

%---------------------------------------------------------------------
\begin{table*}
  \caption{Relevant parameters in LV tests for
    PSRs~B1913+16~\cite{wnt10}, B1534+12~\cite{sttw02},
    B2127+11C~\cite{jcj+06}, and
    J0737$-$3039A~\cite{ksm+06a}. Parenthesized numbers represent the
    uncertainty in the last quoted digits. There is an ambiguity
    between $i$ and $180^\circ-i$; only the value $i < 90^\circ$ is
    tabulated.\label{sm:tab:psr:ecc}}
  \begin{tabular}{lllll}
   \hline\hline
   Pulsar  &  PSR~B1913+16 & PSR~B1534+12 & PSR~B2127+11C &
   PSR~J0737$-$3039A \\
   \hline
   \multicolumn{5}{l}{\tt Observed Quantities} \\
   \hline
   Observational span, $T_{\rm obs}$\,(year) & $\sim25$~\cite{wnt10}
   & $\sim12$~\cite{sttw02} & $\sim12$~\cite{jcj+06} &
   $\sim3$~\cite{ksm+06a} \\
   Right ascension, $\alpha$ (J2000) &
   $19^{\rm h}15^{\rm m}27^{\rm s} \hspace{-1.2mm}.99928(9)$ &
   $15^{\rm h}37^{\rm m}09^{\rm s} \hspace{-1.2mm}.960312(10)$ &
   $21^{\rm h}30^{\rm m}01^{\rm s} \hspace{-1.2mm}.2042(1)$ &
   $07^{\rm h}37^{\rm m}51^{\rm s} \hspace{-1.2mm}.24927(3)$ \\
   Declination, $\delta$ (J2000) &
   $16^\circ06'27^{\prime\prime} \hspace{-1.2mm} .3871(13)$ &
   $11^\circ55'55^{\prime\prime} \hspace{-1.2mm} .5543(2)$ &
   $12^\circ10'38^{\prime\prime} \hspace{-1.2mm} .209(4)$ &
   $-30^\circ39'40^{\prime\prime} \hspace{-1.2mm} .7195(5)$ \\
   Proper motion in $\alpha$, $\mu_\alpha$ (mas\,yr$^{-1}$) &
   $-1.43(13)$ & 1.32(3) & $-1.3(5)$ & $-3.3(4)$ \\
   Proper motion in $\delta$, $\mu_\delta$ (mas\,yr$^{-1}$) & 
   $-0.70(13)$ & $-25.12(5)$ & $-3.3(10)$ & $2.6(5)$ \\
   Spin period, $P$ (ms) & 59.0300032180(5) & 37.9044407982695(4) &
   30.52929614864(1) & 22.699378599624(1) \\
   Orbital period, $P_{\rm b}$ (day) & $0.322997448911(4)$ &
   0.420737299122(10) &
   $0.33528204828(5)$ & $0.10225156248(5)$ \\
   Eccentricity, $e$ & $0.6171334(5)$ & 0.2736775(3) & $0.681395(2)$ &
   $0.0877775(9)$ \\
   Projected semimajor axis, $x$ (lt-s) & $2.341782(3)$ & 3.729464(2) & $2.51845(6)$
   & $1.415032(1)$ \\
   Longitude of periastron, $\omega$ (deg) & $292.54472(6)$ &
   274.57679(5) & $345.3069(5)$ & $87.0331(8)$ \\
   Epoch of periastron, $T_0$ (MJD) & 52144.90097841(4) & 50260.92493075(4) &
   50000.0643452(3) & 53155.9074280(2) \\
   Advance of periastron, $\dot\omega$ (deg\,yr$^{-1}$) &
   $4.226598(5)$ & 1.755789(9) & $4.4644(1)$ & $16.89947(68)$ \\
   Einstein delay parameter, $\gamma$ (ms) & 4.2992(8) & 2.070(2) &
   4.78(4) & 0.3856(26) \\
   Shapiro delay parameter, $s$ & $\cdots$ & 0.975(7) & $\cdots$ &
   $0.99974^{+0.00016}_{-0.00039}$ \\
   Shapiro delay parameter, $r$ ($\mu$s) & $\cdots$ & 6.7(10) &
   $\cdots$ & 6.21(33) \\
   Intrinsic derivative of $P_{\rm b}$, $\dot P_{\rm b}^{\rm int}$
   ($10^{-12}\,{\rm s\,s}^{-1}$) & $-2.396(5)$ & $-0.174(11)$ &
   $-3.95(13)$ & $-1.252(17)$ \\
   Mass ratio, $q\equiv m_1/m_2$ & $\cdots$ & $\cdots$ & $\cdots$ &
   1.0714(11) \\
   \hline
   \multicolumn{5}{l}{\tt Estimated Quantities} \\
   \hline
   Upper limit of $|\dot e|$ ($10^{-14}$\,s$^{-1}$) & 0.22 & 0.27
   & 1.8 & 3.3 \\
   Upper limit of $|\dot x|$ ($10^{-13}$\,${\rm s\,s}^{-1}$) & 0.13
   & 0.18 & 5.5 & 0.37 \\
   \hline
   \multicolumn{5}{l}{\tt Derived Quantities Based on GR} \\
   \hline
   Pulsar mass, $m_1$ (M$_\odot$) & 1.4398(2) & 1.3332(10) & 1.358(10) &
   1.3381(7) \\
   Companion mass, $m_2$ (M$_\odot$) & 1.3886(2) & 1.3452(10) & 1.354(10) &
   1.2489(7) \\
   $\delta X \equiv (m_1 - m_2) / (m_1 + m_2)$ & 0.0181(1) &
   $-0.0045(5)$ & 0.001(5) & 0.0345(4) \\
   Inclination, $i$ (deg) &  47.194(7) & 77.2(1) & 50.1(4) &
   $88.69^{+0.50}_{-0.76}$ \\ 
   Characteristic velocity, ${\cal V}_{\rm O}$ (km\,s$^{-1}$) &
   438.8390(4) & 394.593(1) & 427.426(7) & 625.04(1) \\
\hline
\end{tabular}
\end{table*}
%\end{turnpage}
%---------------------------------------------------------------------

PSR~B1913+16 (a.k.a. PSR~J1915+1606 or the Hulse-Taylor pulsar) was
the first binary pulsar discovered at the Arecibo Observatory. Since
its discovery precision tests of gravity theories in the strong field
were made feasible.  The pulsar has a spin period of 59\,ms and is in
a 7.75-hour orbit. Follow-up radio timing observations verified
Einstein's GR up to a precision of $0.2\%$ (see~\cite{wnt10} and
references therein). The relativistic timing observables for
PSR~B1913+16 include the periastron advance rate $\dot\omega$, the
``Einstein delay'' in terms of the time dilation parameter, $\gamma$,
and the shinkage of the orbit in terms of the time derivative of the
orbital period, $\dot P_b$. The agreement of the predicted quadrupole
damping effects in GR and the measured $\dot P_b$ value constitutes
the first observational evidence of the existence of gravitational
waves.  The timing solution obtained from data taken at Arecibo
telescope from 1981 to 2006~\cite{wnt10} is listed in
Table~\ref{sm:tab:psr:ecc}. During such a time baseline, the periastron
has already rotated out an angle $\sim 100^\circ$, that is the largest
one in our binary samples.

%=====================================================================
\subsubsection{PSR B1534+12}
\label{sm:sec:B1534}
%=====================================================================

PSR~B1534+12 (a.k.a. PSR~J1537+1155) is a 38\,ms pulsar in a 10.1-hour
orbit discovered at the Arecibo Observatory. It has a high orbital
inclination near to $80^\circ$. Besides the
$\dot\omega$-$\gamma$-$\dot P_b$ measurements, it also provided a
measurement of the Shapiro delay that is caused by the curvature of
spacetime near the companion star~\cite{sttw02}. The timing solution
obtained from data taken at the Arecibo telescope for about $12$ years
is listed in Table~\ref{sm:tab:psr:ecc}. The periastron has rotated out
an angle $\sim20^\circ$ during this time baseline. Stairs et
al.~\cite{sttw02} reported an upper limit for $\dot e$, $|\dot e|
\lesssim 3 \times 10^{-15}$, and an upper limit for $\dot x$, $|\dot
x| \lesssim 6.8\times10^{-13}\,{\rm s\,s}^{-1}$, which are valuable
for studies here. Because more than one decade has passed since the
publication of the current timing solution, new data will for sure
provide even more constraining results.

%=====================================================================
\subsubsection{PSR B2127+11C}
\label{sm:sec:B2127}
%=====================================================================

PSR~B2127+11C (a.k.a. PSR~J2129+1210C or M15C) is a pulsar spinning at
a period of 31\,ms in the globular cluster M15 (a.k.a. NGC~7078). It
was discovered at the Arecibo Observatory in 1989. The pulsar has an
8.0-hour orbit, and a companion suspected to be a NS. The Keplerian
orbital parameters of PSR~B2127+11C are nearly identical to those of
PSR~B1913+16 (see Table~\ref{sm:tab:psr:ecc}). Likewise, this pulsar also
provided a $\dot\omega$-$\gamma$-$\dot P_b$ test, and GR passes the
test within $\sim3$\% precision~\cite{jcj+06}.  The most updated
timing solution was obtained from about 12 years of data from the
Arecibo Observatory, with a 5-year gap due to the update of the
telescope~\cite{jcj+06}.

%=====================================================================
\subsubsection{PSR J0737$-$3039A/B}
\label{sm:sec:J0737}
%=====================================================================

PSR~J0737$-$3039A/B (a.k.a. the Double Pulsar) is the first and up to
now the only system that composes two visible pulsars. It was
discovered with the 64-m Parkes raido telescope.  The orbital period
is only 2.5\,hours which makes the binary extremely relativistic with
a characteristic velocity ${\cal V}_{\rm O}\simeq 625\,{\rm
  km\,s}^{-1}$ in GR.  With a measurement of the mass ratio from the
projected semimajor axes of two pulsars, as well the periastron
advance rate, the Einstein delay, the Shapiro delay, and the orbital
shinkage, it provided many tests of gravity theories just from one
system~\cite{ksm+06a}. GR were tested up to a precision of $0.05\%$ by
only using 3 years of data~\cite{ksm+06a}. PSR~J0737$-$3039A is the
recycled pulsar in the binary and is the significantly better timed
one. One notable quantity of its timing parameters is the very large
periastron advance rate, $\dot\omega\simeq17\,{\rm deg\,yr}^{-1}$,
that has rotated the periastron by $\sim50^\circ$ in $3$ years.

\end{document}